# An Investigative Model of Adult Cyberbullying: A Court Case Analysis

*Completed Research Paper*

**Chintha Kaluarachchi**
College of Business and Law
RMIT University
Melbourne, Victoria
s3863295@student.rmit.edu.au

**Darshana Sedera**
Faculty of Business, Arts & Law
Southern Cross University
Gold Coast, Australia
darshana.sedera@gmail.com

**Matthew Warren**
Centre for Cyber Security Research and Innovation (CCSRI)
RMIT University
Melbourne, Victoria
matthew.warren2@rmit.edu.au

## Abstract

*Cyberbullying is a major social issue that is on the rise with a substantial potential to impact a large number of Internet users globally. The growth and rapid proliferation of the Internet and other ubiquitous technologies like social media and smart mobile devices have increased the propensity of cyberbullying, providing it with a wider audience and rapid access. This research developed an investigative model for cyberbullying, specifically developed for adults. Therein, the model considers the cyberbullying journey from conception of the bullying idea, identification of the target to the bullying as an action. The a-priori model is motivated by the General Theory of Crime and the Routine Activity Theory. The a-priori model is then validated using 20 cyberbullying court cases from Australia, Canada, the United States and Scotland.*

**Keywords:** Cyberbullying, General Theory of Crime, Routine Activity Theory, Adults

## Introduction

Cyberbullying among adults has been identified as a widespread, persistent, and serious social issue often cited as a contributor to several high-profile suicides. "Jessica Logan," "Tyler Clementi" and "Brodie Panlock" are some of the notable suicides that have been attributed to cyberbullying. These victims tragically have committed suicide from the emotional trauma they have suffered due to intense, prolonged attacks in cyberspace. Cyberbullying defined as *"an aggressive, intentional act carried out by a group or individual, using electronic forms of contact, repeatedly and overtime against a victim who cannot easily defend him or herself"* (Smith et al. 2008, p. 376), is a global phenomenon. The advent and rapid proliferation of digital technologies such as social media platforms (Leung 2019; Lokuge et al. 2019; Sedera et al. 2016a), ubiquitous mobile devices and connected societies have exacerbated the impact of cyberbullying in recent times (Chan et al. 2019; Sedera and Lokuge 2018).

As a growing social phenomenon, scholars are increasingly calling for a deeper understanding of cyberbullying to formulate policies, procedures, and technologies to alleviate it. Most of the cyberbullying research largely focused on prevalence and predictors of cyberbullying (Lee 2017), comparison of traditional bullying and cyberbullying (Hinduja and Patchin 2010), coping strategies of cyberbullying (Raskauskas and Huynh 2015), the characteristics of cyber perpetrators, cyber victims, and bystanders (Lund and Ross 2017) and risk and protective factors of cyberbullying (Kowalski et al. 2019). While the past literature on cyberbullying has provided a substantial contribution to our current knowledge, there has been considerably less theory-driven research done to study cyberbullying and much of the research on cyberbullying perpetration has been largely atheoretical or lacks a solid theoretical foundation (Barlett 2017; Xiao et al. 2016). Lowry et al. have also shown that *"cyberbullying has not been investigated using a compelling socio-technical theory"* (Lowry et al. 2019, p.1144). This lack of strong theoretical guidance hinders scientific understanding of this phenomenon (Xiao et al. 2016) and limits their application in practice as well (Barlett 2017). Therefore, a theoretically grounded explanation of cyberbullying is desperately needed to help intervention specialists to tailor their curriculum to specifically target this new social phenomenon, that has motivated the present study.

Furthermore, while cyberbullying is prevalent amongst all ages, much of the related research has focused on adolescents or younger populations and the phenomenon among adults is only receiving recent research attention (Jenaro et al. 2018; Xiao et al. 2016). However, it has been recognized that adults too could be victims of cyberbullying (Kowalski et al. 2017; Sedera et al. 2017). Such bullying had taken place in workplaces (Kowalski et al. 2017) and tertiary education settings (Myers and Cowie 2019; Xiao et al. 2016). The ubiquitous nature of email communication has made it an effective tool for cyber aggression, creating unique challenges for organizations (Richard et al. 2020; Sedera et al. 2016b). For example; a 2019 YouGov research survey released in the UK that interviewed 2,034 people shows that almost a quarter (23%) of British adults have experienced cyberbullying and that had led to real-life negative consequences (Independent 2019). A key issue is that most of the cyberbullying education takes place in schools and most of the countries and states have formed policies and laws to protect youth populations, yet adults also equally need protection, and awareness to protect against cyberbullying (Kowalski et al. 2017). The growth of cyberbullying amongst the adult population deemed almost as severe as compared with younger populations (Jenaro et al. 2018) and this has provided the context to this study.

As such, the objective of the present study is to derive a robust and validated cyberbullying model for adults, which allows useful and pragmatic assessment, based on answers derived from the literature and court case analysis. The remainder of the paper will first present a summary of cyberbullying followed by the derivation of the a-priori cyberbullying model employed. Subsequently, the court case study findings are reported, and the paper concludes by summarizing the study contributions, limitations, and future research directions.

## A brief Summary of Cyberbullying

Cyberbullying is a worldwide phenomenon, and it has immerged as a new form of bullying which occurs through the Internet via cell phones, computer devices or handheld devices and can be anonymous and can occur 24 hours a day (Feinberg and Robey 2008). According to the eleven different studies conducted in the United States, the proliferation of cyberbullying has doubled by 18% in 2007 and 36% in 2019 (López-Vizcaíno et al. 2021).

Srivastava et al. (2013) recognized some common types of cyberbullying behaviours such as *"text-based name-calling, use of coarse language, profanity and personal attacks (which may include racist or sexist attacks), flaming (overt attacks), harassment or denigration, cyberstalking, 'outing' of individuals or sending humiliating photos or video messages"* (Srivastava et al. 2013, p.28-29).

Research identifies a range of motivations of cyberbullying:1) seeking revenge (Berger 2007); 2) based on physical appearance, social status and experiences (Tynes et al. 2010); 3) limited social and peer support (Williams and Guerra 2007); 4) high environmental exposure to violence (Calvete et al. 2010); 5) technology capabilities and activities (Kaluarachchi et al. 2020a; Sedera and Lokuge 2020b); and 6) power imbalance (Berger 2007). Previous studies have claimed that a significant portion of college students are perpetrators and/or victims of cyberbullying (Doane et al. 2014). Furthermore, studies have revealed that there are significant bullying patterns, especially among adults in workplaces (Aboujaoude et al. 2015). According to a recent study, nearly half (46.2%) of trainee doctors have experienced cyberbullying in their workplaces, and that has negatively affected their job satisfaction (Lowry et al. 2016). A US-based study conducted with over 3600 adults, found that over 20% of respondents had cyberbullying victimization in their adulthood (Kowalski et al. 2017). The same study found that most of the cyberbullies were colleagues or co-workers (72.7%) followed by a spouse or significant other (69%) or friends (65%) (Kowalski et al. 2017).

Social media sites and mobile devices are especially fertile ground for cyberbullying (Kaluarachchi et al. 2020b), but such behaviours occur in a wide range of online venues (Kowalski et al. 2014; Sedera and Lokuge 2020a). Cyberbullying can take place through online chat and messaging services, text messages, emails, message boards and online forums that allow people to publicly comment other than social media sites (eSafety Commissioner 2020). Previous research has shown that the majority of cyberbullying victimizations among adults (70%) occurred via social networking sites, 50% via instance massages, 49% via emails, and 44% via cellular phones. (Kowalski et al. 2017). Researchers also show that cyberbullying can create consequences for both the bully and the victim (Peebles 2014). Many studies noted that cyberbullying has been linked to negative outcomes at both the psychological and physical levels such as depression (Perren et al. 2010), distress (Sahin 2012), anxiety (Kowalski et al. 2014), increased psychosomatic symptoms (Sourander et al. 2010) and intrusive thoughts of self-harming and suicide ideation (Kowalski et al. 2014).

## Deriving the a-priori Cyberbullying Model

Considering that cyberbullying is a multidisciplinary problem that includes social, psychological, and behavioral aspects, warranted us to employ a transdisciplinary approach for this research. This approach makes this research unique because many disciplines (psychology, information systems, health, and criminology) are integrated to embark on a new theoretical framework to investigate adult cyberbullying perpetration.

First, an a-priori cyberbullying model was derived from the literature. The a-priori model was then adapted and extended through a court cases study entailing 20 court cases from Australia, Canada, the United States and Scotland. The multidisciplinary nature of this research study deserves attention from a multiple lens perspective because the phenomena can be better explained using different theoretical approaches. Researchers have also shown that integrating competing theories is a good approach to provide a strong research contribution (Lowry et al. 2019). Integration of the theories into a single new conceptual model requires important considerations to ensure that the integration is feasible. We consider the two dimensions by Okhuysen and Bonardi (2011) that describe the relationship between the theoretical lenses that are combined. Those two dimensions are 1) proximity of the theoretical lenses that authors seek to combine and 2) the compatibility of their underlying assumptions (Okhuysen and Bonardi 2011). Similarly, the General Theory of Crime (GTC) and Routine Activity Theory (RAT) share a common space, as both theories used in the field of criminology and used to explain how and why individuals engage in crime or deviant behaviours. As mentioned by Okhuysen and Bonardi (2011) the combination, of GTC and RAT, might help to improve the complexity of our theory to fulfil the research challenges.

They also have shown that boundaries or gaps in one theory can be complemented by the other theory (Okhuysen and Bonardi 2011). The Strength of GTC is that it includes individual characteristics that are crucial to criminal or deviant behaviour, however, GTC has not measured the effectiveness of guardianships as a protective factor against crime or delinquent behaviours. Yet, RAT measured the effectiveness of guardianships as a protective factor against crime or delinquent behaviours. However, RAT omitted a clear explanation of individual characteristics that made individuals vulnerable to motivated offenders. Comparing the two, the strength of GTC is in explaining how an individual reacts psychologically to the crime, while the strength of RAT is the inclusion of guardianships to safeguard the victim. Combining GTC and RAT allows us to complement the gaps of each of these theories, making their integration particularly useful.

## *General Theory of Crime (GTC)*

The General Theory of Crime is one of the most cited criminological theories in the empirical literature (Gottfredson and Hirschi 1990). Gottfredson and Hirschi (1990)'s General Theory of Crime offers an overarching theoretical scaffold for the development of a cyberbullying model, given the characteristics that bullying and criminal acts have in common. The GTC argues that delinquency is a manifestation of low self-control; people who have low self-control are more likely to engage in crime or deviant behaviours when presented with the opportunity to do so. Hence, researchers have explicitly examined the relationship between low self-control and the opportunity to commit a crime to explain bullying and cyberbullying behaviours (Jaeyong and Kruis 2020; Lianos and McGrath 2018; Lowry et al. 2019). They found that the combination of low self-control and opportunity has a similar effect on cyberbullying perpetration as it does on crime and impulsive behaviours. The cyberbullying literature also contains various conceptualizations and definitions of the opportunity construct.

## *Routine Activity Theory (RAT)*

The Routine Activity Theory is also an extensively used criminological theory to explain deviant and crime behaviours (Andresen 2006; Holtfreter et al. 2008). Recently, many IS researchers have also begun to adopt RAT to examine cyber victimization and cyberbullying behaviours (Chan et al. 2019; Kalia and Aleem 2017; Navarro and Jasinski 2012). Routine Activity Theory suggests that crime behaviours likely to occur due to three factors such as the presence of a likely offender, a suitable target, and lack of a capable guardianship (Cohen and Felson 1979). This provides a framework to understand the changes in criminal activities. Bullying is increasingly being identified as a crime in many places (Cornell and Limber 2015), and sharing several characteristics with crime (Starosta 2016), therefore, RAT is also a viable solution to explain cyberbullying.

## *The a-priori Model*

A likely offender refers to an individual who might commit a crime or engage in deviant behaviour for any reason (Cohen and Felson 1979). We were noted that 'likely offender' and 'low self-control' occur in combination, in the interests of parsimony those measures were combined into a single measure. Suitable Target indicates the degree of vulnerability posed by the potential victim or property (Cohen and Felson 1979, Navarro and Jasinski 2012). In this study, we conceptualize a 'suitable target' as an individual towards the bully/offender has got negative attitudes. (e.g., jealousy, disliking, revengefulness, sexual preference). Therefore, we formally conceptualize that 'a likely offender' with low self-control and negative attitudes towards the victim/s, are most likely to engage in cyberbullying, when there is a suitable 'crime opportunity' and in absence of 'capable guardianships. As such 'crime opportunity' and 'capable guardianships' act as the moderators in the relationships between cyberbullying attitudes and cyberbullying. Figure 1 depicts the proposed a-priori cyberbullying framework blending the key constructs of the General Theory of Crime and Routine Activity Theory.

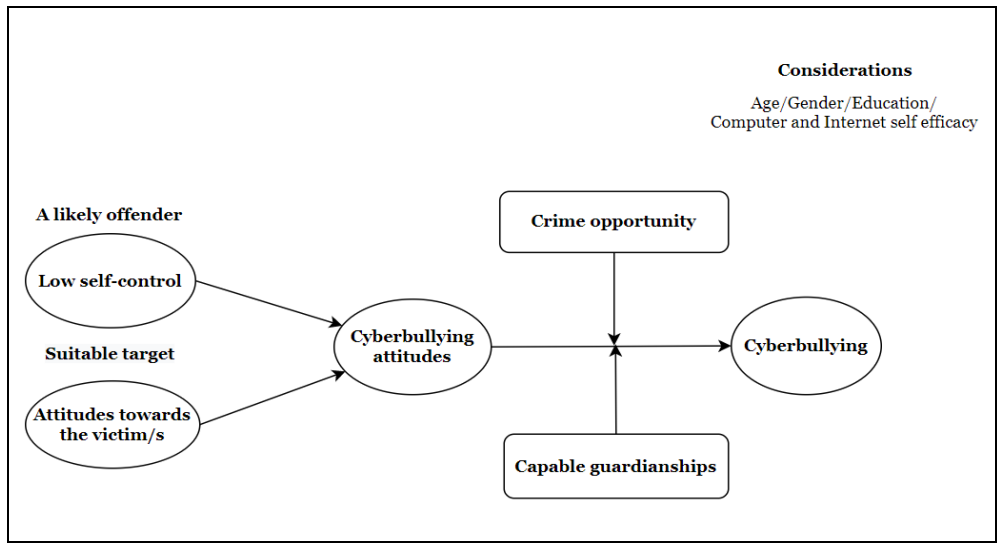

**Figure 1. The a-priori model of adult cyberbullying perpetration**

Five a-priori constructs were identified through this effort and were adopted in this study explained below.

**Low self-control:** We postulate that a likely offender is captured in an individual's low self-control. Low self-control was conceptualized through six different and interrelated dimensions such as impulsivity, preference for simple tasks, preference for physical activities, self-centered tendencies, risk-seeking behaviors, and short temperedness (Gottfredson and Hirschi 1990). Prior literature has shown that individuals with low self-control are at greater risk of committing criminal offences and bullying/cyberbullying (Lianos and McGrath 2018). We proposed that, an offender with low self-control increases cyberbullying attitudes towards cyberbullying perpetration.

**Attitudes towards the victim/s:** Most of the offenders are known to the victim/target such as family members, work colleagues, neighbors, local community members or even friends and may bully or harass someone due to various factors such as thrill-seeking behaviours, sexual preference of the perpetrator, or revengefulness (Kowalski et al. 2014; Vranjes et al. 2017; Walters et al. 2016). Therefore, we posit that high levels of negative attitudes towards the victim/s increase the cyberbullying attitudes and likelihood of cyberbullying.

**Cyberbullying attitudes:** Cyberbullying attitudes have considered a part of one's personality, these attitudes will predict subsequent behaviour and tested with several theoretical frameworks (Barlett 2017). There is a significant direct relationship between cyberbullying attitudes to cyberbullying behaviour (Barlett 2017; Barlett and Kowalewski 2019; Lee and Wu 2018).

**Crime opportunity:** We conceptualize the opportunity as having, four socio-technical factors as potential factors that, if present, may influence the relationship between cyberbullying attitudes and cyberbullying. Those socio-technical factors are 1) time spent online including social media, 2) perceived anonymity, 3) relationships with a vulnerable to deviant or violent peers and 4) cyber or traditional bullying perpetration and victimization.

**Capable guardianships:** According to the routine activity theory, "guardianships are not confined to government officials alone, but rather include anybody whose presence or proximity would discourage a crime from happening" (Baumeister and Boden 1998). We have included offline authorities (e.g., law enforcement) and online mechanisms (e.g., "Facebook/Twitter" help Center, privacy settings, protective software, filters) as capable guardianships that moderate the relationship between cyberbullying attitudes and cyberbullying. We posit that the absence of capable guardianships influences the relationship between cyberbullying attitudes and cyberbullying perpetration.

## Testing the a-priori Cyberbullying Model – Multiple Court Case Analysis

The court case study method was employed to further specify the a-priori cyberbullying model derived through the literature. The court case study method emphasizes qualitative analysis. The qualitative approach address theory construction and theory building, an unlikely quantitative approach that is more concerned with theory testing and verification (Lincoln and Guba 1985). The court case study approach has been chosen because it allows for the capture of knowledge from actual cyberbullies and practitioners involved with the court cases.

The purposive sampling method was used in the selection of the court cases to ensure the appropriateness, purpose, and access to good information related to the research topic. Researcher Patton has shown that *"the validity, meaningfulness and insights generated from qualitative inquiry have more to do with the information-richness of the cases selected and the analytical capabilities of the researcher than with sample size"* (Patton 2002, p. 185). Given the calibration purpose of the study and the information richness of the court cases, the sample size did not have to be large. We used 20 court cases including about 500 pages to validate the a-priori model of cyberbullying.

By following the deductive method, this research begins with theory to develop a coding scheme that is based on certain concepts defined in the theory. Then, the open codes were mapped into the four theoretical paradigms of the GTC and RAT namely 'low self-control', 'attitudes towards the victim/s', 'crime opportunity', 'capable guardianships' and newly introduced construct 'cyberbullying attitudes' based on the attention of each element of the a-priori model has received in the court cases. Next, new constructs of cyberbullying perpetration were sought by inductive reasoning.

Profiling of reviewed literature can provide useful insights about the overall status of the included studies and offers quality assurance, transparency, and the alleviation of selection bias (Gaffar et al. 2015). The analysis of the selected court cases is presented here for this purpose. These court cases were of different countries, different approaches and characteristics but related to cyberbullying. We have used cases from 2008 till 2019 and most of the cases selected related to the year 2016. The profiles of the court cases are 6 cases related to Australia, 7 cases related to Canada, 6 cases related to the United States and 1 court case related to Scotland have been selected for the analysis. Court cases are chosen from a range of socio-demographic conditions; thus, it supports other variables that have been considered. Male and female offenders have been tested, and more often cyberbullying performed by male offenders (about 80%) with an average age of 18 years old or older. The computer and Internet self-efficacy also found significant correlations with cyberbullying. Higher ICT/Internet expertise is associated with a higher level of cyberbullying. Supporting the study objectives, the results indicated that all the a-priori constructs demonstrated strong and significant predictors of adult cyberbullying perpetration. The below section justifies the re-specification of a-priori variables based on the court case findings.

### *Likely offender with low self-control*

The court case analysis demonstrated the overall importance of low self-control as a possible cyberbullying factor. This is consistent with the prior studies (Jaeyong and Kruis 2020; Lianos and McGrath 2018; Lowry et al. 2019). In particular, the sub-constructs of low self-control, such as 'self-centered tendencies' (100%), were highly substantiated with the text of court cases. 'Short temperedness' (60%) makes the second-largest contribution, respectively. In addition, 'preference for simple tasks' (40%) and 'risk-seeking behaviors' (40%) were prominent subconstructs in the analysis. However, the 'preference for physical activities' sub-construct of low self-control did not substantiate through the court cases. Even though this construct is applicable to criminal behaviours, it did not emerge as a significant factor in this analysis. Table 1 record the degree to which the reviewed court cases refer to each element of low self-control.

**Table 1. Summary of reviewed cases in relation to the low self-control**

| | Subconstruct | Total cases | % total | Cases cited |
|---|---|---|---|---|
| **Low Self-Control** | Self-centered tendencies | 20 | 100% | Case #1, Case #2, Case #3, Case #4, Case #5, Case #6, Case #7, Case #8, Case #9, Case #10, Case #11, Case #12, Case #13, Case #14, Case #15, Case #16, Case #17, Case #18, Case #19, Case#20 |
| | Short temperedness | 12 | 60% | Case #1, Case #3, Case #5, Case #7, Case #9, Case #13, Case #14, Case #15, Case #16, Case #17, Case #18, Case# 20 |
| | Preference for simple tasks | 8 | 40% | Case #1, Case #3, Case #5, Case #9, Case #13, Case #14, Case #15, Case #16 |
| | Preference for physical activities | 0 | 0% | |
| | Risk-seeking behaviours | 8 | 40% | Case #2, Case #4, Case #6, Case #8, Case #10, Case #17, Case #19, Case #20, |
| | Depression, anxiety, or other psychological disorders | 4 | 20% | Case #4, Case #6, Case #10, Case #19 |
| | Impulsivity | 3 | 15% | Case #5, Case #11, Case #12, |

Court cases revealed that the offenders preferred short-lived, immediately gratifying, simple, risky, requires little skill tasks, and involve a lack of empathy for the victim and being impulsive. Furthermore, in 20% of the court cases, perpetrators suffered from depression and anxiety as well as having deficits in their psychological functioning. In Case #6, the offender *"revealed high anxiety, moderately high stress and severe depressive symptoms."* Those deficits in their psychological functioning reduce their power or ability to control their emotions, behaviour, or actions and removes the normal hesitations that prevent most individuals from committing these deviant behaviours. This study results show that low self-control is one of the strongest predictors of cyberbullying for women and men, as well as adults as shown in the literature.

### *Suitable Target (Attitudes Towards the Victim/s)*

Another significant factor for cyberbullying perpetration is attitudes towards the victim/s. Consistent with prior cyberbullying studies (Varjas et al. 2010) 'revengefulness' is collectively making the largest contribution (65%) related to 'attitudes towards the victim/s' and the highly substantiated factor with the text of court cases.

The revengefulness was echoed in most of the cases due to factors such as hatred based on dysfunctional relationship (20%), hatred based on work-related conflict (15%) and hatred due to conflicts with neighbors or friends (20%) showing significant relationships. For instance, Case #3 highlighted that the bully took revenge on the victim by posting gruesome Facebook posts and wishing suffering and disease on the victim, concluding, *"I absolutely hate her with every fiber of my being."* Case #10 is another good example, the Barrister Carl Heaton involved with the case, said his client *"who was autistic, was retaliating against people who had been offensive toward him and a Facebook site designed to assist autistic people."*

According to the court case analysis, 'sexual preference' (30%) was the second-greatest reason that led bullies to have degradative attitudes towards the victims, increasing the likelihood of cyberbullying. Court cases reveal that offenders prefer to engage young females in sexual behavior via the Internet due to unusual and deviant sexual interest. The constructs, as derived from court cases, are described in table 2.

**Table 2. Summary of reviewed cases in relation to attitudes towards the victim/s**

| | Subconstruct | Total cases | % total | Cases cited |
|---|---|---|---|---|
| **Attitudes towards the victim/s** | Revengefulness due to work-related conflict. | 3 | 15% | Case #1, Case #7, Case #18 |
| | Revengefulness due to conflicts with neighbors or friends | 4 | 20% | Case #12, Case #14, Case #15, Case #17 |
| | Revengefulness due to dysfunctional relationship | 4 | 20% | Case #5, Case #9, Case #13, Case#20 |
| | Revengefulness due to other reasons | 2 | 10% | Case #3, Case #10, |
| | Sexual preference of the perpetrator | 6 | 30% | Case #2, Case #4, Case #6, Case #8, Case #11, Case #19 |
| | Thrill-seeking | 2 | 10% | Case #4, Case #11 |
| | Disability or transgender aggravation | 1 | 5% | Case #12 |
| | Dislike the victim | 2 | 10% | Case #6, Case #16 |
| | Perceptions of threat | 2 | 10% | Case #1, Case #3 |

The insights in this research further showed a small number of cyberbullies involved with cyberbullying for 'fun' or 'thrill-seeking (10%) without considering its consequences. The offenders in Case #4 indicated that they became involved in the cyberbullying for fun initially saying, *"at the time he and his brother were victimizing the complainant "it was fun, but now feels it was stupid"*, although they realized the danger afterwards. On the other hand, some offenders become involved in cyberbullying simply because they 'do not like the victim' (10%), they have 'perceptions of threat' (10%), their 'transgender aggravation' or the victim has a 'disability' (5%). However, these factors were the least cited factors in the court cases. Overall court cases revealed that 'negative attitudes towards the victim/s' were significant in the context of cyberbullying as shown in the literature that further strengthens their inclusion in the a-priori model.

## *Moderator Variables*

Furthermore, the implied role of crime opportunity as moderating the relationship between low self-control and cyberbullying behaviours was subsequently verified in a handful of studies (Lowry et al. 2019; Starosta 2016). Our study, demonstrating that four measures derived from the literature relate to the notion of opportunity, has parallels in the court case analysis, further strengthening the grounds for their inclusion in the a-priori model.

The problem of cyberbullying continues and has been exacerbated by technological innovations and the widespread dissemination and popularity of social media sites. The advent of the Internet with 24-hour connectivity and social networking sites creates new opportunities for cyberbullying and echoed in all the cases (100%) as mentioned in the prior studies (Barlett et al. 2019; Kowalski et al. 2019; Lianos and McGrath 2018; Lowry et al. 2016).

Besides, technology-enabled perceived anonymity (35%) provides copious opportunities for cyberbullying and substantiated in the court case analysis. Many cases undoubtedly indicated 'perceive anonymity' is a strong factor. The country court involved with Case #15 highlighted that *"Unlike traditional bullying, which usually takes place by a face-to-face encounter, the defendant used the advantages of the Internet to attack his victims from a safe distance, 24 hours a day, while cloaked in anonymity."* These court case findings proved that the technology-enabled perceived anonymity also provide a substantial contribution to cyberbullying.

Scholars have highlighted that individuals with low self-control were more likely to engage in anti-social behaviours when their peers or family members involved in anti-social behaviours (Guo 2016; Smith et al. 2008). The importance of this factor regarding cyberbullying was highlighted in the court case analysis, accounting for 30% of the impact. Table 3 illustrate the degree to which the reviewed court cases refer to each element in the crime opportunity included in the a-priori model (Figure 1) and/or new constructs/relationships derived from the inductive exercise.

**Table 3. Summary of reviewed cases in relation to crime opportunity**

| | Subconstruct | Total cases | % total | Cases cited |
|---|---|---|---|---|
| **Crime Opportunity** | Time spent online including social media | 20 | 100% | Case #1, Case #2, Case #3, Case #4, Case #5, Case #6, Case #7, Case #8, Case #9, Case #10, Case #11, Case #12, Case #13, Case #14, Case #15, Case #16, Case #17, Case #18, Case #19, Case#20 |
| | Perceived anonymity | 7 | 35% | Case #2, Case #6, Case #8, Case #10, Case #15, Case #17, Case #19 |
| | Involvement with previous offences | 3 | 15% | Case #4, Case #10, Case #18 |
| | Traditional bullying or cyberbullying victimization or perpetration | 2 | 10% | Case #6, Case #10 |
| | Involvement with deviant or violent friends or family | 6 | 30% | Case #11, Case #13, Case #14, Case #16, Case #17, Case#20 |
| | Victim's risky ICT use | 4 | 20% | Case #2, Case #4, Case #6, Case #8 |
| | Victim's forced compliance | 4 | 20% | Case #2, Case #4, Case #6, Case #19 |

Consistent with prior studies (Chen et al. 2017; Guo 2016) traditional bullying or cyberbullying victimization or perpetration (10%) also increases the chances of being a cyberbully based on our court case analysis. The court case study has shown that involvement with previous offences (15%) also had the power to influence the relationship between low self-control and cyberbullying.

Other than those literature cited factors, two new subconstructs emerged through the court case analysis and those are 'victim's risky ICT use' (20%) and 'victim's forced compliance' (20%) which show yet another significant contribution of opportunity construct to influence cyberbullying perpetration. Study findings showed a significant correlation between engaging in 'risky ICT use online' and being a cyber victim. Risky ICT behaviours include communicating in social networks with unknown people or share information online with strangers. 'Victim's forced compliance' to do what offenders demand also increases their vulnerability as a suitable target and creates new opportunities for cyberbullying.

'Capable guardianships' is another moderating variable included in the a-priori model and validated via the court case analysis as shown in table 4.

**Table 4. Summary of reviewed cases in relation to capable guardianships**

| | Subconstruct | Total cases | % total | Cases cited |
|---|---|---|---|---|
| **Capable Guardianships** | Online mechanisms | 5 | 25% | Case #1, Case #2, Case #6, Case #9, Case #18 |
| | Offline (Law enforcement) | 17 | 85% | Case #1, Case #2, Case #3, Case #4, Case #5, Case #6, Case #7, Case #8, Case #9, Case #10, Case #11, Case #13, Case #14, Case #15, Case #16, Case #18, Case #19 |
| | Offline (Parents, collage, or another capable guardian) | 7 | 35% | Case #2, Case #3, Case #4, Case #6, Case #11, Case #16, Case#20 |

Offline guardianships such as the country's law enforcement and regulatory system (85%), parents, collage representatives, or other capable guardians (35%) were the highly substantiated types of guardianships with the text of court cases. For instance, Case #4 shows that *"The degree of participation of the accused in the offence was intense, prolonged and only stopped due to the intervention of the* *victim's parents and law enforcement."* An appropriate sentence is essential to prevent cybercrimes including cyberbullying behaviours. However, this study received weak support for online mechanisms (25%), because the existence of online mechanisms provides a tool for users to reduce their exposure as a vulnerable victim, but once cyberbullying commences these mechanisms seems not strong enough to cease the cyberbullying behaviours as offline guardianship.

Each county should have laws that are designed to protect people from cyberbullying and the absence of capable guardianships would encourage such conducts. There are several cases (Case #12, #17 and #20) that victims tragically, have committed suicide from the emotional trauma they have suffered due to intense, prolonged attacks and there were no capable guardianships to save them. Case #20 is a serious example of a lack of capable guardianships. Even though the victim complained to her school authorities, she did not receive help from them and instead offenders escalate the level of harassment because the victim reported about them. This lack of capable guardianships forced her to commit suicide leaving no other choice. By considering these shortcomings, countries and states should appreciate the need for strong laws, regulatory bodies, and trained staff to reduce cyberbullying among adults. Scholars also encouraged this by revealing most of the cyberbullying education takes place in schools and many countries/states have formed policies and laws to protect the young population, yet adults also equally need protection, and awareness to protect against cyberbullying (Kowalski et al. 2017).

### *Dependent Variables (Cyberbullying attitudes to behaviour)*

Consistent with prior studies (Barlett 2017; Barlett and Kowalewski 2019; Lee and Wu 2018) several of the analyzed court cases indicate that most of the offenders (90%) had pro-cyberbullying attitudes which were directly related to cyberbullying perpetration.

Most of the cyberbullying cases demonstrated a persistent, repeated, dense and serious pattern of cyberstalking, harassment and sexual offending which intimidated victims and showed the perpetrators' callous and reckless disregard for their emotional and physical well-being. Also, these cyberbullying behaviours showed a pattern of repetition and escalating seriousness. These cases involved a deliberate posting of offensive material to web pages including social media sites, processing, and distributing child pornography materials and sexual violence against children and adult victims, as well as intimidation and threatening of the victims via the Internet.

## The Overall Re-Specified Cyberbullying Model

The paper shows that parsimoniously integrating two competing theories, GTC and RAT, to explain cyberbullying perpetration among adults. The development and validation of the a-priori model involved (i) consolidation or separation of constructs (ii) introduction of new constructs and measures, and (iii) elimination or revisiting the relevance of the constructs identified in the literature. The constructs, subconstructs and measures, as derived from the literature and court cases are arranged in the conceptual model as shown in figure 2. Where citations did not map into an existing a-priori construct, new construct was created and added to the a-priori model. A new subconstruct of low self-control, 'offender's psychological issues' has been identified and included in the model. Moreover, three new crime opportunities were identified and included in the model: 1) victim's risky ICT use; 2) victim's forced compliance; and 3) offender's previous offences.

Subsequent to mapping, if any construct was not substantiated, it was removed. Therefore 'preference for physical activities' has been removed from the low self-control construct due to its non-relevance to the cyberbullying context. Capable guardianship divided into two categories 1) online mechanisms and 2) offline guardianships (law enforcement) because these two subconstructs received mixed support from the court case analysis. Figure 2 illustrates the validated cyberbullying model which collectively represents all the relevant aspects of the cyberbullying perpetration among adults.

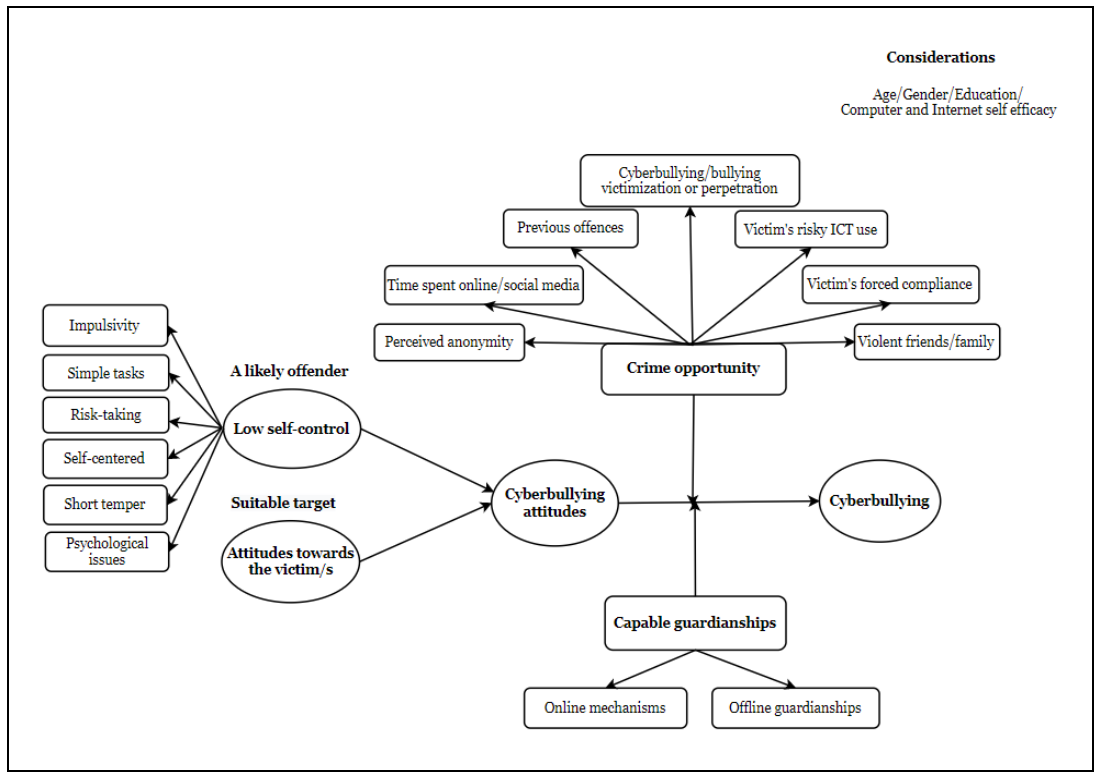

**Figure 2. The overall re-specified cyberbullying model**

While we recognized all these subconstructs are important, each subconstruct received mixed support from the court case analysis. For instance, 'time spent online including social media' is the most dominant and highly substantiated opportunity with the text of court cases. However, 'bullying or cyberbullying victimization or perpetration' only mentioned twice in the court cases. Therefore, future researchers may use these subconstructs based on their research context and questions.

## Conclusion and Contributions

The phenomenon of cyberbullying has become a major problem among adults, and it creates persistent psychological problems. Generally, the cyberbullying literature was found to be a lack of an overarching framework to guide interventions that combat cyberbullying and assist adults to stay safe online. This paper proposes an investigative model for cyberbullying that follows the process of cyberbullying from initial idea to identification of the target, to the bullying action, considering any salient moderators. The model is inspired by two well-established criminological theories that had not been combined in this way before. Theoretically, this research contributes to the information systems (IS) literature by exploring cyberbullying perpetration among adults. This study shows the value of integrating competing theories in a holistic and parsimonious manner to explain cyberbullying among adults. The integration of theories reveals that the combined theoretical framework explains more variance in cyberbullying than each of the theories individually. Overall, this research is expected to stimulate the development of evidence-based policy positions and interventions that combat cyberbullying and assist adults to stay safe online.

## Limitations and Future Research

This research mainly had two limitations. Firstly, the research presented in this paper has been limited by the small number of studies conducted relating to adult cyberbullying. Secondly, although great care was taken to review the court cases thoroughly some limitations recorded. This study only reviewed court cases related to Australia, Canada, Europe, and the United States. We did not find court case transcripts related to Asian and African countries. As future work, another analysis could be conducted to review cases related to those continentals to see if our cyberbullying constructs and relationships remain the same for those countries as well.

## Acknowledgements

This research was supported by the Australian Government Research Training Program Scholarship and RMIT University CCSRI.